\documentclass[12pt,preprint]{aastex}

\begin{document}

\title{The Nucleus of Main-Belt Comet 259P/Garradd
\footnote{Based on observations obtained at the Southern Astrophysical Research (SOAR) telescope, which is a joint project of the Minist\'{e}rio da Ci\^{e}ncia, Tecnologia, e Inova\c{c}\~{a}o (MCTI) da Rep\'{u}blica Federativa do Brasil, the U.S. National Optical Astronomy Observatory (NOAO), the University of North Carolina at Chapel Hill (UNC), and Michigan State University (MSU).}
}
\author{Eric M.\ MacLennan$^{a,b,c}$ \& Henry H.\ Hsieh$^{a,d,e}$}

\affil{$^a$ Institute for Astronomy, Univ.\ of Hawaii, 2680 Woodlawn Drive, Honolulu, HI 96822\newline
       $^b$ Dept.\ of Physics and Astronomy, Northern Arizona University, Flagstaff, AZ 86001\newline
       $^c$ Dept. of Earth and Planetary Sciences, University of Tennessee, Knoxville, TN 37996\newline
       $^d$ Hubble Fellow\newline
       $^e$ Visiting astronomer, Cerro Tololo Inter-American Observatory, National Optical Astronomy Observatory}
\email{emaclenn@utk.edu, hsieh@ifa.hawaii.edu}

\begin{abstract}
We present observations of the main-belt comet 259P/Garradd, previously known as P/2008 R1 (Garradd), obtained in 2011 and 2012 using the Gemini North Telescope on Mauna Kea in Hawaii, and the SOAR Telescope at Cerro Pachon in Chile, with the goal of computing the object's phase function and nucleus size.  We find an absolute magnitude of $H_R=19.71\pm0.05$ mag and slope parameter of $G_R=-0.08\pm0.05$ for the inactive nucleus, corresponding to an effective nucleus radius of $r_e=0.30\pm0.02$ km, assuming an $R$-band albedo of $p_R=0.05$.  We also revisit observations reported for 259P while it was active in 2008 to quantify the dust mass loss and compare the object with other known main-belt comets.
\end{abstract}


\section{Introduction}
Main-belt comets (MBCs) orbit completely within the main asteroid belt, yet exhibit cometary activity indicative of the sublimation of volatile ice \citep{hsi06}.  By providing evidence that ice exists in the present-day asteroid belt, they present an invaluable opportunity to evaluate hypotheses that icy main-belt objects may have played a significant role in the primordial delivery of terrestrial water \citep[e.g.,][]{mor00}.

Comet 259P/Garradd (hereafter, 259P) was the fourth MBC to be discovered \citep{gar08}, and has a semimajor axis of $a=2.726$ AU, an eccentricity of $e=0.342$, and an inclination of $i=15.90\degr$.  When it was active in 2008, \citet{jew09} estimated that it had a dust production rate on the order of $10^{-2}$~kg~s$^{-1}$ and an upper limit CN production rate of $Q_{\rm CN}<1.4\times10^{23}$ s$^{-1}$, and determined that it was dynamically unstable on timescales of $\sim$20-30 Myr.  This dynamical result indicated that 259P is likely not native to its current location, and probably originated elsewhere in the asteroid belt.  An origin in the outer solar system, however, could not be ruled out.

\citet{jew09} computed an upper limit to the effective radius of the nucleus of $r_e=0.7$ km, but since the object remained active throughout their observational campaign, they were unable to ascertain the true nucleus size.  In this letter, we present observations aimed at measuring this key physical parameter.

\section{Observations\label{section_observations}}
Observations of 259P were obtained using the 8.1~m Gemini North telescope (Programs GN-2011A-Q-15, GN-2011B-Q-17, and GN-2012A-Q-68) on Mauna Kea in Hawaii and the 4.2 m Southern Astrophysical Research (SOAR) telescope (Program 2012A-0461) on Cerro Pachon in Chile.  Gemini observations were made using the imaging mode of the Gemini Multi-Object Spectrograph \citep[GMOS; image scale of 0$\farcs$1454 pixel$^{-1}$;][]{hoo04} and a Sloan Digital Sky Survey (SDSS) $r'$ filter.  SOAR observations were made using the SOAR Optical Imager \citep[SOI; 0$\farcs$154 pixel$^{-1}$;][]{sch04} and a Kron-Cousins $R$-band filter. Non-sidereal tracking of the object was performed for all observations.

\begin{figure}
\plotone{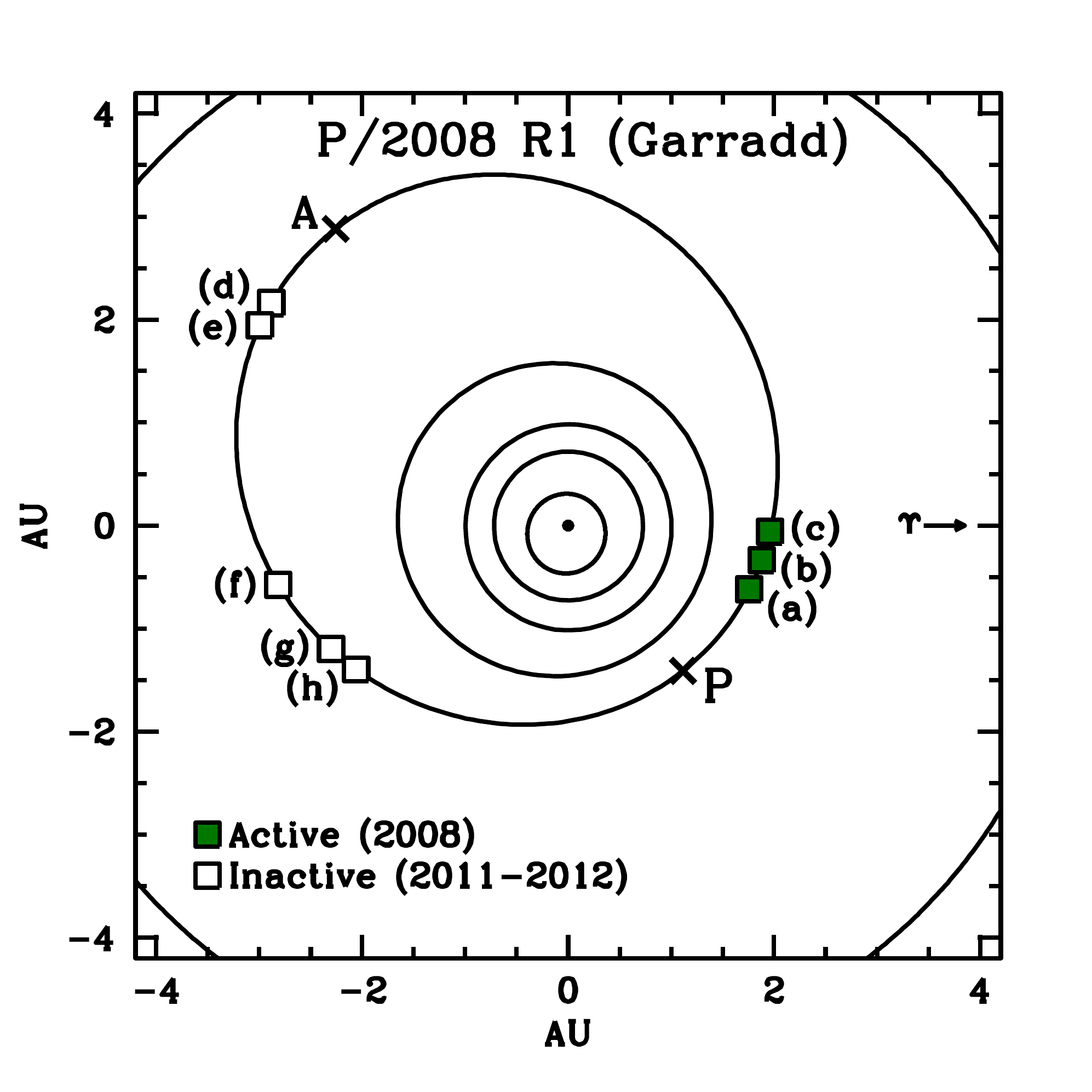}
\caption{Orbital position plot for 259P observations described in Table \ref{obslog} and Table \ref{photexcess}. The Sun is shown at the center as a solid dot, with orbits of Mercury, Venus, Earth, Mars, 259P, and Jupiter (from the center outward) shown as black lines. Green squares mark positions where 259P was observed to be active in 2008 by \citet{jew09} and open squares mark inactive positions in 2011 and 2012. The perihelion (P) and aphelion (A) positions are marked with crosses. Observations plotted are from (a) 2008 September 26 - October 03, (b) 2008 October 22, (c) 2008 Nov 11, (d) 2011 February 28 - March 11, (e) 2011 March 26 - April 01, (f) 2012 January 21 - February 01, (g) 2012 April 15, and (h) 2012 May 13-15, where (a)-(c) are from \citet{jew09} and (d)-(h) are from this work.}
\label{fig_orbitplot}
\end{figure}

\begin{figure}
\plotone{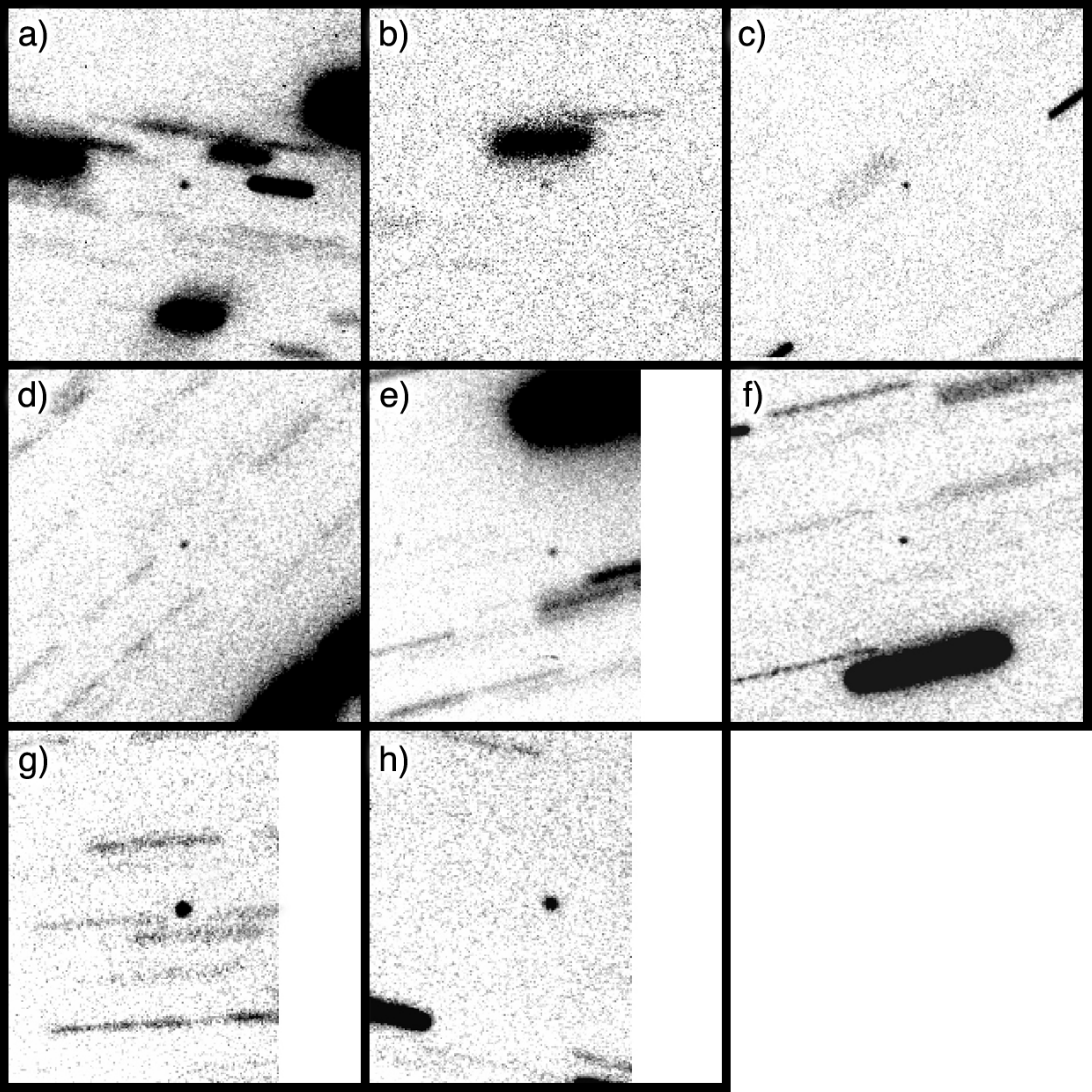}
\caption{Composite R-band images of 259P constructed from observations obtained on (a) 2011 February 28 (2400 s of effective exposure time), (b) 2011 March 11 (1200 s), (c) 2011 March 31 (1800 s), (d) 2011 April 1 (3600 s), (e) 2012 January 21 (2700 s), (f) 2012 January 31 (1980 s), (g) 2012 April 15 (1800 s), and (h) 2012 May 15 (1800 s). All panels are constructed from data obtained with Gemini North. 259P is shown at the center of each panel measuring 30$\arcsec$ by 30$\arcsec$ in size, with North up and East to the left.}
\label{fig_images}
\end{figure}

We obtained our observations on twelve nights between UT 2011 February 28 and 2012 May 15 during a portion of the object's orbit when active dust emission was expected to be minimal or absent (Table~\ref{table_obslog}; Figure~\ref{fig_orbitplot}).  Inactivity was confirmed by summing the images taken on each night and visually inspecting the resulting composite images (Figure~\ref{fig_images}), ensuring that these observations reflect the properties of the inactive nucleus.

Standard image calibration (bias subtraction and flat-field reduction) was performed for all images. Flat fields were constructed from dithered images of the twilight sky.  Object and field star photometry was performed using circular apertures for which optimum sizes were chosen depending on the amount of field-star trailing present due to the non-sidereal tracking of the object, and nightly seeing conditions.  Absolute calibration of object photometry was then performed using field star magnitudes from the SDSS catalog \citep{yor00}. Transformation to the $BVRI$ system was done for all Gemini measurements, assuming solar colors for the object, allowing our phase function to be computed for $R$-band.

\section{Results\label{section_results}}
\subsection{Phase Function Analysis\label{section_phasefunction}}
An object's brightness variation with solar phase angle (its phase function) is dependent on the surface properties of the object, such as regolith porosity, albedo, particle size distributions and roughness \citep{hel89,mui02}.  Thus, an understanding of the phase functions for 259P and other objects of interest, such as other MBCs and other asteroids and comets, provide a means for comparing their surface properties and discerning possible contrasts or similarities. Phase function parameters also help us accurately determine the expected magnitude of the object at any observing geometry, assuming it is inactive.

To determine the phase function of 259P, we first normalize our measured apparent $R$-band magnitudes, $m(R,\Delta,\alpha)$, to $R=\Delta=1$ AU using
\begin{equation}
m(1,1,\alpha)=m(R,\Delta,\alpha)-5\log R\Delta
\end{equation}
The resulting reduced magnitude, $m(1,1,\alpha)$, depends only on the phase angle and rotational phase of the object, where the rotational properties of 259P are presently unknown.  
All of our Gemini data were taken at unknown rotational phases and thus represent ``snapshot'' observations.  In our more extensive SOAR data set which covered a longer time baseline, we observe a photometric range (peak to trough) of $\Delta m=0.6\pm0.1$~mag.  Unfortunately, we were unable to derive a plausible rotational lightcurve from these data, and so the rotational period and amplitude of 259P remain unknown.  However, we see a range in brightnesses over our Gemini observations that is consistent with the magnitude range from our SOAR data, suggesting that 259P's photometric range is in fact approximately $\Delta m\sim0.6$~mag.  In our analysis below, we assume that by using sparse sampling, we can compensate for rotational effects, and include such variations as a source of uncertainty in our final phase function solution.

Fitting our data to the IAU $H,G$ phase function \citep{bow89}, we find best-fit parameters of $H_R=19.71\pm0.05$~mag and $G_R=-0.08\pm0.05$.  For reference, this corresponds to linear phase function parameters of $m_R(1,1,0)=20.08 \pm 0.05$ mag and $\beta=0.050\pm0.005$~mag~deg$^{-1}$.  For comparison, the slope parameters of MBCs 133P, 176P, and 238P are $G=0.04\pm0.05$ \citep{hsi09b}, $G=0.26\pm0.05$ \citep{hsi09b}, and $G=-0.03\pm0.10$ \citep{hsi11b}, respectively. The range of slope parameters for kilometer-scale Themis asteroids, of which 176P and 133P are identified as members, is $-0.23<G<0.60$ \citep{hsi08}. These parameters suggest that 259P is similar in composition to other MBCs, which resemble dark, B-type asteroids \citep{hsi09b,lic11}, and Themis family asteroids which are mostly C-type and B-type asteroids \citep{flo99,tho89,laz04}.

Using a geometric $R$-band albedo of $p_R = 0.05$ \citep[measured for other MBCs;][]{hsi09b}, we estimate an effective nucleus radius of $r_e\sim0.3$~km using
\begin{equation}
p_R r_e^2 = (2.24 \times 10^{16}) \times 10^{0.4[m_\odot - H_R]}
\end{equation}
where $m_\odot = -27.07$ is the apparent solar $R$-band magnitude \citep{har80,har82,har90}, and $r_e$ is in km.

\subsection{Analysis of 2008 Activity\label{section_dustactivity}}
Using the phase function we have derived for 259P, we can estimate the amount of dust that was present in the 2008 observations obtained by \citet{jew09}. We calculate reduced $R$-band magnitudes from their reported photometry and compare them with nucleus magnitudes that we predict from our derived phase function to determine the ratio of scattering surfaces of dust, $A_d$, to that of the nucleus $A_N$ using
\begin{equation}
\frac{A_d}{A_N}=\frac{1-10^{0.4(m_{tot}(1,1,\alpha) - m_R)}}{10^{0.4(m_{tot}(1,1,\alpha) - m_R)}}
\end{equation}
where $m_{tot}$ is the calculated reduced magnitude from \citet{jew09} when the total flux of 259P was measured, including any dust surrounding the nucleus, and $m_R$ is the expected reduced magnitude according to our derived IAU phase function. Using the estimated nucleus radius for 259P from Section~\ref{section_phasefunction}, a total dust mass, $M_d$, can be estimated assuming dust grains with radii of $a=10~\mu$m and bulk densities of $\rho=1300$~kg~m$^{-1}$ \citep{hsi04} using
\begin{equation}
M_d = \case{4}{3}\rho a\cdot A_N\left(\frac{A_d}{A_N}\right)
\end{equation}
Dust to nucleus scattering surface ratios and dust mass values derived from the reported photometry from \citet{jew09} are shown in Table~\ref{photexcess}. Fitting a linear function to our computed dust masses over this time period, we find a dust dissipation rate from the coma of $\sim10^{-2}$~kg~s$^{-1}$ (Figure~\ref{fig_phasecurve}), consistent with the rate estimated by \citet{jew09}.


\section{Discussion}
\subsection{Nature of 259P's Activity}
The discovery of disrupted asteroids, i.e., asteroids where comet-like activity is directly caused by an impact event and not by sublimation of volatile ices, has raised questions as to whether we can assume that the activity of any of the previously known MBCs is sublimation-driven \citep[e.g.,][]{jew12}.  \citet{hsi12a} considered a variety of observable criteria that could potentially be used to distinguish between MBCs that exhibit sublimation-driven activity and disrupted asteroids.  These criteria include long-lived activity, constant or increasing coma dust mass, dust morphology, and recurrent activity.

\begin{figure}
\plotone{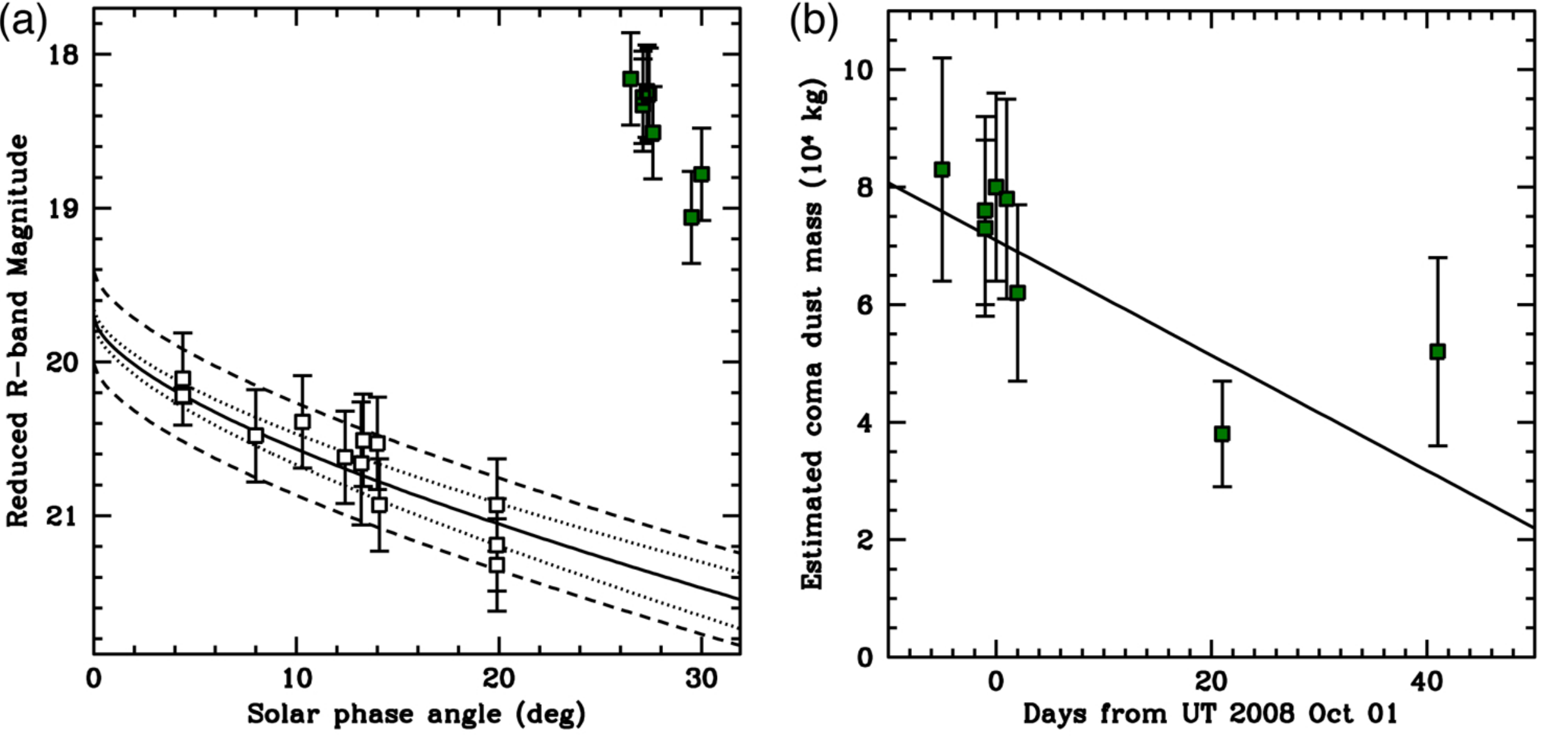}
\caption{(a) Best-fit IAU phase function (solid line) for 259P, where dotted lines show the range of uncertainty due to phase function parameter uncertainties, and dashed lines show the expected photometric range due to rotational variation, assuming $\Delta m=0.6$~mag. Open squares mark photometry from this work (Table \ref{obslog}) when 259P is inactive and green squares indicate photometry from \citet{jew09} (Table \ref{photexcess}), in which 259P displayed cometary activity. (b) Estimated coma dust masses (green squares; from Table~\ref{photexcess}) of 259P during its 2008 active period, and a best-fit linear mass loss function (solid line), plotted versus time \citep[after][]{jew09}.}
\label{fig_phasecurve}
\end{figure}

Considering the first two criteria, we note that 259P's activity was observed over a total period of only 49 days and that it faded by $\sim$60\% over that period (Figure~\ref{fig_phasecurve}).  For comparison, the impact-produced dust cloud of (596) Scheila was observed for approximately one month, but faded by 30\% in just 8 days at the end of that period \citep{jew11}{bod11}.  By contrast, the coma mass of MBC P/2010 R2 (La Sagra) increased by $\sim$110\% over 4.5 months \citep{hsi12c}, while the coma mass of MBC P/2006 VW$_{139}$ remained constant for $\sim$30 days \citep{hsi12b}.  The photometric behavior of the latter two objects suggests the action of prolonged dust production, consistent with sublimation and inconsistent with impulsive ejection.  While a decrease in activity strength by itself is not a definitive indicator of the source of that activity, the fact that 259P's activity was only observed as it was fading leaves open the possibility that it was generated by an impulsive event like an impact, and not prolonged sublimation.

Numerical dust modeling was used to infer the cometary nature of the activity of MBCs 133P/Elst-Pizarro, 176P/LINEAR, 238P/Read, P/La Sagra, and P/2006 VW$_{139}$ \citep{hsi04,hsi09a,hsi11a,hsi12a,hsi12c,mor11}, and also to infer the impact-driven nature of the activity of P/2010 A2 (LINEAR) and (596) Scheila \citep{jew10,sno10,ish11}, but has never been performed for 259P.  Given the myriad assumptions and simplifications used in such models, however, any results would necessarily need to be considered inconclusive pending confirmation via other analyses \citep[cf.][]{jew12}, so the lack of a modeling analysis of 259P is not a critical deficiency.

The strongest observable criterion identified to date for establishing cometary activity is recurrent activity with intervening periods of inactivity \citep{hsi12a,jew12}.  It is simple to reconcile such behavior with a sublimation-driven process, where activity can be modulated either by periodically varying solar heating over an object's orbit (as for classical comets), or by seasonally varying solar illumination of an isolated active site \citep[such as the site of an impact where subsurface ice has been excavated and exposed;][]{hsi04,hsi09,cap12}.  In contrast, recurrent activity cannot be easily explained by other dust ejection mechanisms except in extremely contrived scenarios \citep{jew12}.

In the case of 259P, we have not yet had the opportunity to test for the recurrence of activity as it is just now about to complete one full orbit since its discovery in 2008.
The next opportunity to observe 259P after it passes perihelion on 2013 January 25 is in July 2013, at which point it will be at a true anomaly of $\nu\sim60\degr$.  In 2008, activity was observed between $\nu=29.2\degr$ and $\nu=48.6\degr$.  Therefore, unfortunately, it will not be possible to compare activity levels from 2008 and 2013 over the same orbital arcs.  Nonetheless, since we have now measured a precise nucleus size, we can simply search for photometric excesses (beyond the expected brightness of a bare nucleus) in 2013 data, even if no visible cometary features (e.g., a tail or coma) are detected, and immediately compute the mass of any coma that may be present (subject to rotational phase uncertainties).  Then by extrapolating the fading trend observed in 2008 (Table~\ref{table_2008photom}; Figure~\ref{fig_phasecurve}) to the orbital arc covered by 2013 observations, we will be able to ascertain whether there are significant changes in activity level between the two epochs.

To date, recurrent activity has only been confirmed for 133P and 238P \citep{hsi04,hsi10,hsi11b}.  Confirming recurrent activity for 259P would strongly indicate that its activity is sublimation-driven, while a lack of recurrent activity would strongly suggest that 259P's 2008 activity could have been generated by an impact.  As such, observations beginning in July 2013 to confirm either the presence or absence of activity are critical for determining this object's true nature.

\subsection{Comparison to Other MBCs}
Effective nucleus radii have now been measured (or estimated using assumed albedos) for five MBCs: 133P \citep[$r_e=1.9$~km;][]{hsi09b,hsi10,bau12}, 176P \citep[$r_e=2.0$~km;][]{hsi09b,hsi11a,bau12}, 238P \citep[$r_e=0.4$~km;][]{hsi11b}, P/La Sagra \citep[$r_e=0.7$~km;][]{hsi12c}, and now, 259P ($r_e=0.3$~km; this work).  While it is still too early to draw firm conclusions about the size distribution of MBCs based on measurements of just five objects, it is nonetheless interesting to note that all are relatively small (a few km across, or smaller), and three of the five have effective nucleus radii of $<$1~km.

This apparent preference for small nuclei may be due to their lower escape velocities.  Given the low $\sim1-3$~m~s$^{-1}$ ejection speeds of dust particles for some MBCs \citep{hsi04,hsi09a}, where the escape velocities of the largest known MBCs are $\sim$1~m~s$^{-1}$, it is possible that similar emission on larger objects may simply be unobservable because ejected particles never actually escape the body to produce visible activity.

Another potential explanation for the small sizes of MBC nuclei is related to the likely ages of these objects.  The collisional destruction timescale of an object varies with its size, with object diameters of 10~km, 1~km, and 0.1~km corresponding to collisional destruction timescales of $\sim$1-10~Gyr, $\sim$0.1-1~Gyr, and $\sim$10-100~Myr, respectively \citep{che04,bot05}.  Thus small objects that still exist today are likely to have formed more recently (e.g., in a fragmentation or cratering event involving a larger parent body) than larger ones.  Based on their very small nucleus sizes, 238P, 259P, and P/La Sagra could therefore be much younger than the age of the solar system.  This implication is intriguing considering independent dynamical findings that 133P and P/2006 VW$_{139}$ may be products of extremely recent ($<$10~Myr) asteroid family-forming fragmentation events \citep{nes08,nov12}.

While thermal models suggest that water ice could persist for Gyrs from hundreds to just a few meters under the surface of a main-belt asteroid \citep[e.g.,][]{sch08,pri09}, ice could be present much closer to the surface in a much younger asteroid, rendering that ice far more susceptible to ``activation'' via collisional excavation \citet[cf.][]{hsi04}.  Extremely young asteroids would also be able to retain near-surface ice for wider ranges of thermal inertias, obliquities, and latitudes than those found by \citet{sch08}.  As more MBCs are discovered, it will be important to continue measuring nucleus sizes to improve our understanding of their distribution and determine whether this apparent preference for small MBC nucleus sizes is real.

In terms of activity strength relative to nucleus size (as parameterized by $A_d/A_N$), 259P's level of activity is similar to that of 238P \citep{hsi11b}, and $\sim$1-2 orders of magnitude higher than those of 133P and 176P \citep{hsi12c}.  However, given 259P's small nucleus size, in absolute terms, the total scattering cross-section of dust in the coma, $A_d$, is actually comparable to those of 133P and 176P, as well as that of 238P \citep{hsi12c}.  While there are many complicating factors, such as different escape velocities for the wide range of MBC nucleus sizes and the fact that the amount of dust resident in the coma at a given time does not necessarily correspond to the instantaneous dust production rate, the observation that absolute activity levels are comparable among MBCs (at least within an order of magnitude) suggests that they may have comparably-sized active areas.  If the collisional activation model described by \citet{hsi04} is correct, this could suggest that the MBCs observed to date were activated by similarly-sized impactors \citep[$\sim$1~m;][]{hsi09}, though many more observations of many more MBCs will be needed to confirm this possible trend.

As discussed above, we were unable to determine 259P's rotational period from our SOAR data, which is unfortunate because a comparison to other MBCs would have been useful.  The only MBCs with measured rotation rates --- 133P \citep[$P_{\rm rot}=3.471$~hr;][]{hsi04} and 176P \citep[$P_{\rm rot}=22.23$~hr;][]{hsi11a} --- both exhibit extreme rotation (one relatively fast and one relatively slow) compared to other asteroids \citep[cf.][]{mas09}.  Our precise determinations of the size and phase function of 259P's nucleus have now made it possible to make much more reliable predictions of its brightness though, which are of obvious benefit to future observation planning.  As such, attempts to measure 259P's rotation period at the next appropriate opportunity (i.e., when the object is expected to be inactive, but also close enough to the Earth and Sun to be bright enough to obtain high-quality photometric data) are highly encouraged.

\section*{Acknowledgements}
E.M.M.\ was supported by the NSF grant 0757887 through its Research Experience for Undergraduate program hosted at the University of Hawaii's Institute for Astronomy.
H.H.H.\ is supported by NASA through Hubble Fellowship grant HF-51274.01 awarded by the Space Telescope Science Institute, which is operated by the Association of Universities for Research in Astronomy (AURA), Inc., for NASA, under contract NAS 5-26555.
We thank Chad Trujillo, Andrew Stephens, and numerous queue observers at Gemini, and Sean Points, Patricio Ugarte, and Alberto Pasten at SOAR for their assistance in acquiring observations, as well as Marco Micheli for assisting with ephemeris predictions, and David E.\ Trilling for useful discussion and providing access to \LaTeX\ software to E.M.M.

The Gemini Observatory is operated by the Association of Universities for Research in Astronomy, Inc., under a cooperative agreement with the NSF on behalf of the Gemini partnership.
The Cerro Tololo Inter-American Observatory is a part of the National Optical Astronomy Observatory which is operated by AURA, under contract with the National Science Foundation.
SDSS-III (http://www.sdss3.org/) is funded by the Alfred P.\ Sloan Foundation, the Participating Institutions, NSF, and the U.S.\ Department of Energy Office of Science, and managed by the Astrophysical Research Consortium for the SDSS-III Collaboration.

\begin{deluxetable}{lcrrrccrcc}
\tablewidth{0pt}
\tablecaption{Observation Log and Photometric Results\label{obslog}}
\tablecolumns{10}
\tablehead{
\colhead{UT Date} &
\colhead{Tel.\tablenotemark{a}} &
\colhead{N\tablenotemark{b}} &
\colhead{$t_{\rm exp}$\tablenotemark{c}} &
\colhead{$\nu$\tablenotemark{d}} &
\colhead{$R$\tablenotemark{e}} &
\colhead{$\Delta$\tablenotemark{f}} &
\colhead{$\alpha$\tablenotemark{g}} &
\colhead{$m_R(R,\Delta,\alpha)$\tablenotemark{h}} &
\colhead{$m_R(1,1,\alpha)$\tablenotemark{i}}}
\startdata
2011 Feb 28 & Gemini &  4 &  2400 & 194.5 & 3.60 & 2.71 &  8.0 & $25.4\pm0.1$   & $20.5\pm0.3$ \\
2011 Mar 11 & Gemini &  2 &  1200 & 195.8 & 3.59 & 2.77 & 10.3 & $25.4\pm0.1$   & $20.4\pm0.3$ \\
2011 Mar 26 & Gemini &  2 &  1800 & 197.6 & 3.57 & 2.90 & 13.2 & $25.7\pm0.2$   & $20.7\pm0.4$ \\
2011 Mar 31 & Gemini &  2 &  1800 & 198.2 & 3.57 & 2.95 & 14.0 & $25.6\pm0.1$   & $20.5\pm0.3$ \\
2011 Apr 01 & Gemini &  4 &  3600 & 198.3 & 3.57 & 2.96 & 14.1 & $26.0\pm0.1$   & $20.9\pm0.3$ \\
2012 Jan 21 & Gemini &  3 &  2700 & 240.7 & 2.89 & 2.72 & 19.9 & $25.4\pm0.1$   & $20.9\pm0.3$ \\
2012 Jan 31 & Gemini & 11 &  1980 & 242.5 & 2.86 & 2.55 & 19.9 & $25.5\pm0.1$   & $21.2\pm0.3$ \\
2012 Feb 01 & Gemini &  5 &   900 & 242.7 & 2.86 & 2.53 & 19.9 & $25.6\pm0.1$   & $21.3\pm0.3$ \\
2012 Apr 15 & SOAR   & 22 & 13200 & 258.0 & 2.59 & 1.60 &  4.4 & $23.35\pm0.05$ & $20.25\pm0.05$ \\
2012 Apr 15 & Gemini &  6 &  1800 & 258.0 & 2.59 & 1.60 &  4.4 & $23.2\pm0.1$   & $20.1\pm0.1$ \\
2012 May 13 & Gemini &  3 &   540 & 264.6 & 2.49 & 1.57 & 12.4 & $23.6\pm0.1$   & $20.6\pm0.3$ \\
2012 May 15 & Gemini &  6 &  1800 & 265.1 & 2.48 & 1.58 & 13.3 & $23.5\pm0.1$   & $20.5\pm0.3$
\enddata
\tablenotetext{a}{Telescope}
\tablenotetext{b}{Number of images}
\tablenotetext{c}{Total effective exposure time, in seconds}
\tablenotetext{d}{True anomaly, in degrees}
\tablenotetext{e}{Heliocentric distance, in AU}
\tablenotetext{f}{Geocentric distance, in AU}
\tablenotetext{g}{Solar phase angle (Sun-259P-Earth)}
\tablenotetext{h}{Observed mean R-band magnitude}
\tablenotetext{i}{Estimated reduced R-band magnitude (normalized to $R$\ =\ $\Delta=1$ AU) of object at mid-point of full photometric range (assumed to be 0.60~mag) of rotational lightcurve}
\label{table_obslog}
\end{deluxetable}

\begin{deluxetable}{lcrrrrcccc}
\tabletypesize{\footnotesize}
\tablewidth{0pt}
\tablecaption{Analysis of 2008 Photometry\label{photexcess}}
\tablecolumns{10}
\tablehead{
\colhead{UT Date} &
\colhead{$m_{tot}(R,\Delta,\alpha)$\tablenotemark{a}} &
\colhead{$\nu$\tablenotemark{b}} &
\colhead{$R$\tablenotemark{c}} &
\colhead{$\Delta$\tablenotemark{d}} &
\colhead{$\alpha$\tablenotemark{e}} &
\colhead{$m_{tot}(1,1,\alpha)$\tablenotemark{f}} &
\colhead{$m_R(1,1,\alpha)$\tablenotemark{g}} &
\colhead{${A_d}/{A_N}$\tablenotemark{h}} &
\colhead{$M_d$\tablenotemark{i}}}
\startdata
2008 Sep 26 & $19.69\pm0.05$ & 29.2 & 1.85 & 1.09 & 26.5 & $18.16\pm0.05$ & $21.33\pm0.18$ & $17.5\pm4.0$ & $8.3\pm1.9$ \\
2008 Sep 30 & $19.88\pm0.03$ & 30.9 & 1.86 & 1.13 & 27.1 & $18.28\pm0.03$ & $21.35\pm0.18$ & $16.0\pm3.4$ & $7.6\pm1.6$ \\
2008 Sep 30 & $19.93\pm0.03$ & 30.9 & 1.86 & 1.13 & 27.1 & $18.33\pm0.03$ & $21.35\pm0.18$ & $15.3\pm3.2$ & $7.3\pm1.5$ \\
2008 Oct 1  & $19.86\pm0.02$ & 31.4 & 1.86 & 1.13 & 27.3 & $18.24\pm0.02$ & $21.36\pm0.18$ & $16.8\pm3.3$ & $8.0\pm1.6$ \\
2008 Oct 2  & $19.90\pm0.03$ & 31.8 & 1.87 & 1.14 & 27.4 & $18.26\pm0.03$ & $21.37\pm0.18$ & $16.5\pm3.5$ & $7.8\pm1.7$ \\
2008 Oct 3  & $20.17\pm0.06$ & 32.3 & 1.86 & 1.15 & 27.6 & $18.51\pm0.06$ & $21.37\pm0.18$ & $13.0\pm3.2$ & $6.2\pm1.5$ \\
2008 Oct 22 & $21.09\pm0.04$ & 40.4 & 1.91 & 1.33 & 29.5 & $19.06\pm0.04$ & $21.45\pm0.19$ & $8.0\pm2.0$ & $3.8\pm0.9$ \\
2008 Nov 11 & $21.21\pm0.10$ & 48.6 & 1.97 & 1.56 & 30.0 & $18.78\pm0.10$ & $21.47\pm0.19$ & $10.9\pm3.4$ & $5.2\pm1.6$
\enddata
\tablenotetext{a}{Observed magnitude of 259P from \citet{jew09}} 
\tablenotetext{b}{True anomaly, in degrees}
\tablenotetext{c}{Heliocentric distance, in AU}
\tablenotetext{d}{Geocentric distance, in AU}
\tablenotetext{e}{Solar phase angle (Sun-259P-Earth), in degrees}
\tablenotetext{f}{Total reduced magnitude (at $R=\Delta=1$~AU) of 259P}
\tablenotetext{g}{Expected reduced magnitude from IAU phase law}
\tablenotetext{h}{Inferred ratio of scattering surface area of dust to scattering surface area of the nucleus}
\tablenotetext{i}{Estimated dust mass within photometric aperture, in $10^4$ kg}
\label{table_2008photom}
\end{deluxetable}

\end{document}